\documentclass[doublecol]{epl2} 

\usepackage{amsfonts}
\usepackage{amsmath}
\usepackage{hyperref}
\hypersetup{colorlinks=true,linkcolor=blue,citecolor=blue,filecolor=blue,urlcolor=blue,breaklinks=true}

\title{Comment on ``Relativistic quantum oscillator model under the effects of the violation of Lorentz symmetry by an arbitrary fixed vector field'' by Faizuddin Ahmed}
\shorttitle{Comment on ``RQO model under the effects of the VLS by an arbitrary fixed vector field''} 
\author{Andrés G. Jirón Vicente\inst{1} \and Luis B. Castro\inst{2} \and Angel E. Obispo\inst{3,4}}
\shortauthor{A.G. Jirón Vicente \and L.B. Castro \and A.E. Obispo }
\institute{
\inst{1} Facultad de Ciencias Naturales y Matemáticas, Universidad Nacional del Callao, 07001, Bellavista, Callao, Perú\\
  \inst{2} Departamento de F\'{\i}sica, Universidade Federal do Maranh\~{a}o,
  65080-805, S\~{a}o Lu\'{\i}s, MA, Brazil\\ \email{lrb.castro@ufma.br, luis.castro@pq.cnpq.br}\\
  \inst{3} Universidad Tecnológica del Perú (UTP), Lima, Perú\\
  \inst{4} Departamento de Ciencias, Universidad Privada del Norte (UPN), Lima, Perú\\
}

\abstract{We obtain the correct expressions for the energy and normalized eigenfunctions for a spin-zero relativistic quantum oscillator model under the violation of Lorentz symmetry defined by an arbitrary constant vector field $v^{\mu}$.}

\begin{document}

\maketitle

In a recent paper in this Journal, F. Ahmed \cite{EPL138:20001:2022} has studied a spin-zero relativistic quantum oscillator model under the violation of Lorentz symmetry defined by an arbitrary constant vector field $v^{\mu}$. The purpose of this comment is point to out a misleading treatment on the prescription used by the KG oscillator (quantization condition) and the normalization condition, which have lead to a wrong expressions for the energy spectrum and normalized wave function. We use this opportunity to obtain the correct solutions to this problem in a more transparent way.

In this context, the relativistic quantum dynamics of a scalar particle under
Lorentz symmetry breaking effects is described by the equation (1) in Ref. \cite{EPL138:20001:2022}, where the interaction known as the Klein-Gordon (KG) oscillator is achieved by means the substitution (prescription of Mirza et al) \cite{CTP42:664:2004}
\begin{equation}\label{substi}
\vec{p}\cdot\vec{p}\rightarrow\left(\vec{p}+iM\omega\vec{r}\right)\cdot\left(\vec{p}-iM\omega\vec{r}\right)\,.
\end{equation}
\noindent Thereby, the Klein-Gordon oscillator in cylindrical coordinates $(t,r,\varphi,z)$ is given by
\begin{equation}
\left[-\frac{\partial^2}{\partial t^2}+\nabla^2-M^2\omega^2r^2+2M\omega-g(v^\mu \partial_\mu)^2 -M^2\right]\Psi=0, \label{osc2}
\end{equation} 
\noindent where $\nabla^2$ is the Laplacian operator in cylindrical system. The differential equation in Ref.~\cite{EPL138:20001:2022} [Eq. (3)] is different to our result (\ref{osc2}) due to it was obtained from a wrong prescription $\vec{p}\cdot\vec{p}\rightarrow\left(\vec{p}-iM\omega\vec{r}\right)\cdot\left(\vec{p}+iM\omega\vec{r}\right)$. The same mistake is considered in \cite{EPL136:61001:2022}.    

The conservation law for $J^{\mu}$ follows from the standard procedure and it results in $\partial_{\mu}J^{\mu}=0$, where
\begin{equation}\label{quadriC}
J^{\mu}=-\frac{1}{M}\mathrm{Im}\left[ \Psi^{\ast}\left( \partial^{\mu}\Psi+gv^{\mu}v^{\nu}\partial_{\nu}\Psi\right)\right]\,.
\end{equation}
\noindent From this, the normalization condition can be expressed as $\int d\tau J^{0}=\pm 1$, where the plus (minus) sign must be used for a positive (negative) charge. Note that the normalization condition directly depends on the choice of configuration of the vector field $v^{\mu}$, very different from the one used in Ref. \cite{EPL138:20001:2022} (Schr\"{o}dinger-like normalization).
 
Following the procedure of Ref. \cite{EPL138:20001:2022}, we will discuss three different configurations of the vector field $v^\mu$.

\subsection{Case A: Background four-vector with the configuration $v^\mu = (v^0,\vec{0})$}

In this scenario, $v^0 =a$ is a constant and $\vec{0}$ is a $3D$ zero vector. In this case, the Eq. (\ref{osc2}) is rewritten as 
\begin{equation}\label{osc3}
\left[-(1+ga^2)\frac{\partial^2}{\partial t^2}+\nabla^2-M^2\omega^2r^2+2M\omega-M^2\right]\Psi=0. 
\end{equation}
\noindent Meanwhile,
\begin{equation}\label{J0_A}
J^{0}=-\frac{(1+a^{2}g)}{M}\mathrm{Im}\left( \Psi^{\ast}\partial_{t}\Psi \right)\,.
\end{equation}
\noindent Now, in order to solve this equation analytically, we adopt the usual decomposition
\begin{equation}\label{ansatz}
\Psi(t,r,\varphi,z)=\frac{\psi(r)}{\sqrt{r}}\mathrm{e}^{-iEt+il\varphi+ikz}\,. 
\end{equation}
\noindent Substituting (\ref{ansatz}) into (\ref{osc3}), the radial function $\psi(r)$ obeys the radial equation
\begin{equation}\label{eqefA}
\left[ \frac{d^{2}}{dr^{2}}+\Lambda- M^{2}\omega^{2}r^{2}-\frac{(l^{2}-\frac{1}{4})}{r^{2}}\right]\psi(r)=0,
\end{equation}
\noindent where
\begin{equation}
\Lambda=(1+ga^2)E^2-M^2-k^2+2M\omega\,.
\end{equation}
\noindent Following the standard procedure, the quantization condition furnishes
\begin{equation}\label{enerA}
E_{n}=\pm\sqrt{\frac{1}{1+a^{2}g}\left( M^{2}+k^{2}+2M\omega n \right)}\,,
\end{equation}
\noindent where $n=2n_{r}+\vert l\vert=0,1,2,\ldots$ is the principal quantum number, with $n_r$ being a nonnegative integer. The solution for all $r$ can be written as
\begin{equation}\label{solnorA}
\psi(r)=N_{n}r^{\vert l\vert+\frac{1}{2}}e^{-M\omega r^2/2}L^{(\vert l\vert)}_{\frac{n-\vert l\vert}{2}}(M\omega r^{2})\,,
\end{equation}
\noindent where $N_{n}$ is a normalization constant. By using the charge density (\ref{J0_A}), together with the normalization condition, with $\vert E\vert\neq 0$, one can determine the normalization constant  
\begin{equation}\label{consA}
N_{n}=\sqrt{\frac{2M^{\vert l\vert+2}\omega^{\vert l\vert+1}\Gamma(\frac{n-\vert l\vert}{2}+1)}{(1+a^{2}g)\vert E_{n}\vert\,\Gamma(\vert l\vert+\frac{n-\vert l\vert}{2}+1)}}\,,
\end{equation}
\noindent where $E_{n}$ is given by (\ref{enerA}).

\subsection{Case B: Background four-vector with the configuration $v^\mu = (0,\vec{v})$}

In this scenario, $v^\mu =(0,0,0,v^z)$, where $v^z =c$ is a constant. Then, considering this case, the eq. (\ref{osc2}) is rewritten as 
\begin{equation}\label{osc4}
\left[-\frac{\partial^2}{\partial t^2}+\nabla^2-gc^2\frac{\partial^2}{\partial z^2}-M^2\omega^2r^2+2M\omega -M^2\right]\Psi=0\,. 
\end{equation}
\noindent Meanwhile,
\begin{equation}\label{J0_B}
J^{0}=-\frac{1}{M}\mathrm{Im}\left( \Psi^{\ast}\partial_{t}\Psi \right)\,.
\end{equation}
\noindent In this case, the quantization condition furnishes
\begin{equation}\label{enerB}
E_{n}=\pm\sqrt{M^{2}+\left(1-gc^{2}\right)k^{2}+2M\omega n }\,,
\end{equation}
\noindent with $n=0,1,2,\ldots$. The solution for all $r$ is given by (\ref{solnorA}) and, in this case, the normalization constant with $\vert E\vert\neq0$ can be written as
\begin{equation}\label{consB}
N_{n}=\sqrt{\frac{2M^{\vert l\vert+2}\omega^{\vert l\vert+1}\Gamma(\frac{n-\vert l\vert}{2}+1)}{\vert E_{n}\vert\,\Gamma(\vert l\vert+\frac{n-\vert l\vert}{2}+1)}}\,,
\end{equation}
\noindent where $E_{n}$ is given by (\ref{enerB}).

\subsection{Case C: Background four-vector with the configuration $v^\mu = (v^{0},\vec{v})$}

In this scenario, we choose $v^\mu =(v^0,0,0,v^z)$, where $v^0 =a$ and $v^z =c$ are constants. Then, considering this case, the Eq. (\ref{osc2}) is rewritten as 
\begin{equation}\label{osc5}
\begin{split}
\left[-(1+ga^2)\frac{\partial^2}{\partial t^2}+\nabla^2-gc^2\frac{\partial^2}{\partial z^2}-2agc\frac{\partial^2}{\partial t \partial z}\right.\\
\left.-M^2\omega^2r^2+2M\omega -M^2\right]\Psi=0\,. 
\end{split}
\end{equation}
\noindent This last expression is very different to the result obtained in Ref.~\cite{EPL138:20001:2022} [Eq. (18)] due to it was obtained from a misleading treatment of the term $(v^{0}\partial_{0}+v^{j}\partial_{j})^{2}$. Meanwhile,
\begin{equation}\label{J0_C}
J^{0}=-\frac{1}{M}\mathrm{Im}\left\{ \Psi^{\ast}\left[\left(1+a^{2}g\right)\partial_{t}\Psi+acgk\partial_{z}\Psi\right]\right\}\,.
\end{equation}
\noindent In this case, the quantization condition furnishes
\begin{equation}\label{enerC}
E_{n}= \frac{acgk}{1+ga^2} \pm \sqrt {\left(\frac{acgk}{1+ga^2}\right)^2 +\frac{M^2+(1-gc^2)k^2+2M\omega n}{(1+ga^2)}}\,,
\end{equation}
\noindent with $n=0,1,2,\ldots$. Once again, the solution for all $r$ is given by (\ref{solnorA}) and, in this case, the normalization constant with $\vert (1+a^{2}g)E-acgk\vert\neq 0$ can be written as
\begin{equation}\label{consC}
N_{n}=\sqrt{\frac{2M^{\vert l\vert+2}\omega^{\vert l\vert+1}\Gamma(\frac{n-\vert l\vert}{2}+1)}{\vert (1+a^{2}g)E_{n}-acgk\vert\Gamma(\vert l\vert+\frac{n-\vert l\vert}{2}+1)}}\,,
\end{equation}
\noindent where $E_{n}$ is given by (\ref{enerC}).

Finally, we summarize the main conclusions of this comment. Firstly, using the correct prescription [see Eq. (\ref{substi})], we compute exact expressions for the energy spectrum of this system. In all cases, the expressions for the energy spectrum and the second-order differential equations in our manuscript represent a corrected version of those shown in \cite{EPL138:20001:2022}, particularly in the case $C$, where, in addition, some algebraic errors were reported. Secondly, we show that the normalization condition directly depends on the choice of configuration of the vector field $v^\mu$, very different from the one used in Ref. \cite{EPL138:20001:2022} (Schr\"{o}dinger-like normalization). The correctly normalized eigenfunctions were also presented as part of the results of our work.

\acknowledgments
This work was supported in part by means of funds provided by CNPq, Brazil, Grant No. 311925/2020-0 (PQ), FAPEMA, Brazil, Grant No. UNIVERSAL-01220/18, and CAPES, Brazil.

\bibliographystyle{eplbib}


\end{document}